\documentclass[12pt,]{iopart}
\usepackage{graphicx}
\usepackage{iopams}

\begin{document}

\title{Robust Coherent Superposition of States using Quasiadiabatic Inverse Engineering}

\author{Yen-Huang Liu and Shuo-Yen Tseng}

\address{Department of Photonics, National Cheng Kung University, No. 1 University Rd., Tainan 70101, Taiwan}
\ead{tsengsy@mail.ncku.edu.tw}
\vspace{10pt}
\begin{indented}
\item[]August 2017
\end{indented}

\begin{abstract}
We use the invariant-based inverse engineering subject to the quasiadiabatic condition to produce robust and high fidelity coherent superposition of quantum states.
The inverse engineering provides shortcuts to the desired quantum-state evolution while the quasiadiabaticity provides robustness with respect to errors.
We derive simple pulses with low areas which are robust with respect to pulse area and detuning.
\end{abstract}

\vspace{2pc}
\noindent{\it Keywords}: quantum control, adiabatic techniques, quantum superpositions
%\submitto{\JPB}
%\maketitle

\section{INTRODUCTION}
The coherent manipulation of quantum systems with time-dependent fields is a fundamental problem in quantum information processing and atomic/molecular physics,
with many applications in atom interferometry, metrology, chemical interaction control, nuclear magnetic resonance, quantum information processing, etc. \cite{allen87,bergmann98,vitanov01,kral07,saffmann10,guerin03}
An important class of coherent manipulation involves the production of robust superpositions of states \cite{unanyan04,sangouard04,daems05}.
For two-level systems, there are several techniques for coherent manipulation of the states of a quantum system, for example, $\pi$-pulses \cite{allen87}, composite pulses \cite{tycko83,ivanov11,torosov11,genov13}, optimal control theory (OCT) techniques \cite{garon13,boscain02}, and adiabatic techniques \cite{vitanov01,guerin03,yatsenko04}.
In general, $\pi$-pulses are fast but highly sensitive to variations in pulse parameters.
The composite pulses are robust, exact, but slow; and require accurate control of pulse phase and intensity.
The OCT method is fast and efficient, but the optimization is complicated because it requires a large number of parameters.
Adiabatic techniques provide robustness but require a long time, and do not always lead to the exact target state.

Since the pioneer works on adiabatic optimization by Demirplak and Rice \cite{demirplak03,demirplak05}, a number of techniques have been developed to speed up adiabatic evolutions.
Shortcut to adaibaticity (STA) has been proposed as a set of techniques to speed up the slow adiabatic processes, while keeping or enhancing robustness \cite{chen10,ruschhaupt12,lu13,torrontegui13}.
Also, a similar technique called designer evolution of quantum systems by inverse engineering (DEQSIE) \cite{vitanov15} has also been developed for synthesizing Hamiltonians for the desired quantum state evolution.
In particular, these techniques can generate the exact target state; when combined with perturbative treatment of the errors, it leads to a technique for robust and high-fidelity quantum state control by a single-shot shaped pulse (SSSP) \cite{daems13,ndong15}.
The SSSP technique achieves robustness by nullifying integrals corresponding to different orders of errors in the excitation field profile, providing a smooth pulse that can be viewed as a faster version of the composite pulses.
These approaches, while being robust against particular errors by design, do not guarantee adiabaticity. 
A design that is robust against a particular error might be susceptible to other sources of error. 

A family of processes have been developed to optimize the efficiency and robustness of adiabatic passage using local adiabaticity constraints \cite{guerin02,guerin11,quan10,roland02,sofia15}.
The fast quasiadiabatic dynamics (FAQUAD) approach reduces the time by homogeneously distributing the adiabaticity parameter along the process using a single control parameter \cite{sofia15}. 
However, the process does not always lead to the exact target state.
On the other hand, the invariant-based inverse engineering STA \cite{ruschhaupt12,lu13} allows one to freely design the system evolution from the desired initial state to the final state. 
In this paper, the invariant-based inverse engineering and quasiadiabatic condition are combined for the first time for the production of coherent superposition of states.
This new approach, called quasiadiabatic inverse engineering (QIE), allows precise transfer to the target state and designs the system evolution under the quasiadiabatic condition such that the adiabaticity parameter along the process remains a constant. 
Processes designed with this new approach are exact and now robust against various sources of errors, instead of only robust against specific errors by design. 

%Put \label in argument of \section for cross-referencing
\section{MODEL}
\subsection{Inverse engineering using dynamical invariants}
We consider a two-level quantum system driven by a time-dependent Hamiltonian of the form
\begin{equation}
\label{H}
H(t)=\frac{\hbar}{2}\left[
\begin{array}{cc}
-\Delta(t) & \Omega(t)\\
\Omega(t) & \Delta(t)
\end{array}
\right],
\end{equation}
in the basis $|1\rangle \equiv$ $1\choose 0$, $|2\rangle \equiv$ $0\choose 1$. 
In a laser-adapted interaction picture under the rotating wave approximation, $\Delta(t)$ and $\Omega(t)$ are the time-dependent detuning and (real) Rabi frequency.
The idea of the invariant based STA is that one can describe the system evolution using the eigenstates of the dynamical invariant $I(t)$ with the invariant satisfying $\frac{\partial}{\partial t} I+\frac{i}{\hbar}[H,I]=0$, and the eigenstates of the invariant are decoupled during system evolution.
An arbitrary solution of the time-dependent Schr\"{o}dinger equation $i\hbar\frac{\partial}{\partial t}|\Psi(t)\rangle=H(t)|\Psi(t)\rangle$ can be written as a superposition of the eigenstates, $|\Psi(t)\rangle=c_0e^{i\kappa_0(t)}|\psi_0(t)\rangle+c_{\bot}e^{i\kappa_{\bot}(t)}|\psi_{\bot}(t)\rangle$, where $c_0$ and $c_{\bot}$ are time-independent amplitudes, and $\kappa_0(t)$ and $\kappa_{\bot}(t)$ are the Lewis-Riesenfeld phases \cite{lewis69}.
We can parameterize the eigenstates of the invariant as
\begin{equation}
\label{phi}
|\psi_0(t)\rangle=            %e^{i \gamma/2} 
\left[\begin{array}{c}
                      \cos \frac \theta 2~e^{{-i\beta}/2} \\
                      \sin \frac\theta 2~e^{{i\beta}/2}
                    \end{array}
                    \right],
\end{equation}
and the orthogonal one (for all times $\langle \psi_0 (t) | \psi_{\bot} (t) \rangle=0$)
\begin{equation}
\label{phiorthogonal}
|\psi_{\bot} (t) \rangle=%e^{-i \gamma/2} 
\left[\begin{array}{c}
                      \sin \frac \theta 2~e^{{-i\beta}/2} \\
                      -\cos\frac \theta 2~e^{{i\beta}/2}
                    \end{array}
                    \right].
\end{equation}
The system evolution is now described by these new parameters $\theta$ and $\beta$. 
We substitute Eqs. (\ref{phi}) or (\ref{phiorthogonal}) directly into the Schr\"{o}dinger equation and obtain the following auxiliary differential equations \cite{ruschhaupt12,lu13,vitanov15,daems13,ndong15}:
\begin{eqnarray}
\label{thetadot}
\dot{\theta}  &=& \Omega \sin\beta, \\
\label{betadot}
\dot{\beta} &=& \Omega \cot\theta \cos\beta+\Delta.
%\label{gammadot}
%\dot{\gamma} &=& \dot{\theta}\cot\beta/\sin\theta,
\end{eqnarray}
These equations are equivalent to those obtained by the invariant dynamical theory \cite{lewis69}, since $|\psi_0(t)\rangle \langle \psi_0(t)|$ ($|\psi_{\bot}(t)\rangle \langle \psi_{\bot}(t)|$) is a dynamical invariant \cite{ruschhaupt12,lu13,daems13}. 

The inverse engineering is achieved by choosing the parameters $\theta(t)$ and $\beta(t)$ first, and then constructing the Hamiltonian (obtaining the corresponding $\Omega(t)$ and $\Delta(t)$) inversely through Eqs. (\ref{thetadot}) and (\ref{betadot}). 
By setting the boundary conditions of $\theta(t)$ at the initial and final times $t_i$ and $t_f$, one can obtain a given target state from an initial state up to a phase factor.
There is still much freedom to design $\theta(t)$ and $\beta(t)$, except for the boundary conditions. 
In Refs. \cite{ruschhaupt12,lu13,daems13,ndong15}, the inverse engineering was used together with perturbation theory calculations to design evolutions that are robust against particular errors.
Here, we apply the quasiadiabatic condition to achieve system robustness.
As we show in the following, the freedom allows one to bring the system to quasiadiabaticity by designing the system evolution.

\begin{figure}[b]
\center
\includegraphics[width=4in]{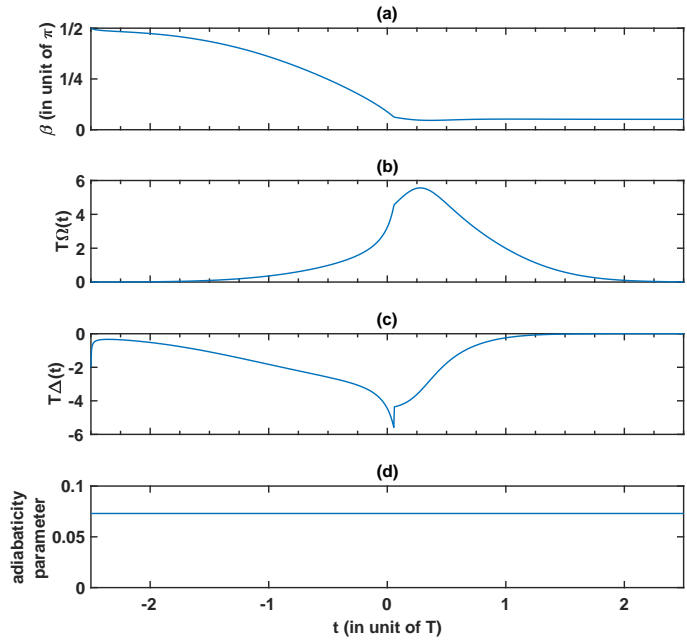}%
\caption{\label{f1}(a) $\beta(t)$ obtained subject to the quasiadiabatic condition with $c=0.073$. (b) The corresponding Rabi frequency $\Omega(t)$ with an area of $1.97\pi$. (c) The corresponding detuning $\Delta(t)$. (d) The calculated adiabaticity parameter of the QIE protocol.}
\end{figure}

\begin{figure}[b]
\center
\includegraphics[width=4in]{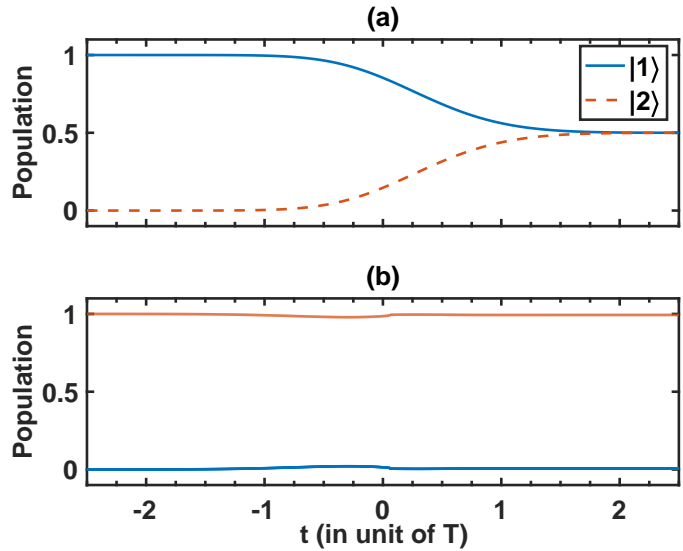}%
\caption{\label{f2} Time evolution of the population in (a) the $|1\rangle$ and $|2\rangle$ basis, and (b) the adiabatic basis.}
\end{figure}

\subsection{Quasiadiabaticity}
By definition, adiabaticity occurs when the state of a quantum system described by a time-dependent Hamiltonian $H(t)$, initially prepared in an instantaneous eigenstate $|\phi_i(0)\rangle$, remains close to the instantaneous eigenstate $|\phi_i(t)\rangle$ during its evolution, as long as $H(t)$ changes slowly.
For a two-level system, the adiabaticity condition reads \cite{schiff81}, $\hbar\left|\frac{\langle\phi_1(t)|\partial_t\phi_2(t)\rangle}{E_1(t)-E_2(t)}\right|\ll1$, where $|\phi_{1,2}(t)\rangle$ is the instantaneous eigenstate with eigenvalue $E_{1,2}(t)$.
Although the adiabatic condition above has shown to be problematic if the change in eigenstate is significant \cite{marzlin04} and a necessary and sufficient condition has been provided in \cite{wang16}, the condition is still applicable in this work because the obtained states vary slowly.
 
While the adiabatic theorem applies the adiabaticity condition on the whole process, 
the FAQUAD approach instead imposes the adiabaticity condition locally in time; that is \cite{sofia15}
\begin{equation}
\label{faquad}
\hbar\left|\frac{\langle\phi_1(t)|\partial_t\phi_2(t)\rangle}{E_1(t)-E_2(t)}\right|=c.
\end{equation}
The spirit of this approach is to distribute the nonadiabatic transitions homogeneously in time so that the constant $c$ is minimized. 
$c$ is solved by imposing consistency with the physically imposed boundary conditions on the Hamiltonian.   
This approach drives the system as close to the adiabatic limit as possible during the duration of transfer as allowed by the physical system (determined by $c$).
Like the other local adiabaticity strategies, the final population in the target state oscillates due to quantum interference, thus high fidelity cannot be guaranteed except at specific process times \cite{sofia15}.

\subsection{Quasiadiabatic inverse engineering (QIE)}
The QIE approach adapts the quasiadiabaticity condition in Eq. (\ref{faquad}) to fix the system adiabaticity along the process at a chosen constant $c$.
The combination of inverse engineering and quasiadiabaticity ensures high fidelity and makes the process as adiabatic as possible at all times.
Solving for the corresponding instantaneous eigenstates of the Hamiltonian in Eq. (\ref{H}) and substituting into Eq. (\ref{faquad}), we obtain the quasiadiabatic condition
\begin{equation}
\label{faquad2}
\left|\frac{\dot{\Omega}(t)\Delta(t)-\Omega(t)\dot{\Delta}(t)}{2(\Omega(t)^2+\Delta(t)^2)^{3/2}}\right|=c.
\end{equation}
Our strategy is then to inverse engineer $\Omega(t)$ and $\Delta(t)$ by designing $\theta(t)$ and $\beta(t)$ subject to the boundary conditions with the quasiadiabaticity constraint in Eq. (\ref{faquad2}).
The system of nonlinear differential equations Eqs. (\ref{thetadot}), (\ref{betadot}) and (\ref{faquad2}) can be solved by numerical integration. 
The key difference between the QIE and FAQUAD, besides the inverse engineering which ensures high fidelity in QIE, is that the quasiadiabaticity constant $c$ in Eq. (\ref{faquad}) in QIE can be chosen arbitrarily small while it is given in FAQUAD.
In QIE, a value for $c$ is set first, with a smaller value meaning a more adiabatic process; then, Eqs. (\ref{thetadot}) and (\ref{betadot}) are integrated under the constraint of Eq. (\ref{faquad2}).
The parameters $\Omega(t)$ and $\Delta(t)$ can then be inversely obtained through Eqs. (\ref{thetadot}) and (\ref{betadot}). 
A smaller $c$ value will lead to a larger pulse area for the obtained $\Omega(t)$, as more energy is needed to drive the system closer to adiabaticity.
Next, we apply this strategy to inverse engineer robust coherent superposition of states in a two-level system.

\section{NUMERICAL RESULTS}
\subsection{Coherent superposition of states}
From Eq. (\ref{phi}), we can see that to describe transfer from the ground state at initial time $t_i$ to the final coherent superposition of the ground and excited states with equal weights at final time $t_f$, up to a phase factor, we should set the following boundary conditions $\theta(t_i)=0$ and $\theta(t_f)=\pi/2$. Following Ref. \cite{ndong15}, we choose the following smooth parameterization for $\theta(t)$ (from $-\infty$ to $+\infty$)
\begin{equation}
\label{theta}
\theta(t)=\frac{\pi}{4}[\textrm{erf}(t/T)+1],
\end{equation}
where $T$ is a characteristic time. 
The target state is then 
\begin{equation}
\label{psitar}
|\Psi_{tar}\rangle=\frac{1}{\sqrt{2}}
\left[\begin{array}{c}
                      e^{{-i\beta(t_f)}/2} \\
                      e^{{i\beta(t_f)}/2}
                    \end{array}
                    \right].
\end{equation}
Knowing $\theta(t_i)=0$ and applying l'H\^{o}pital's rule to Eq. (\ref{betadot}) repeatedly, we find an additional boundary condition $\beta(t_i)=\pi/2$.
In fact, a global phase term $e^{-i\gamma/2}$ can be multiplied to Eq. (\ref{phi}) to make the initial phase arbitrary \cite{daems13}.
We set $\gamma=0$ here for the ease of discussion.
The problem is now reduced to solving for $\beta(t)$ subject to the quasiadiabatic condition Eq. (\ref{faquad2}) with the chosen $c$ value. 
We do not control $\beta(t_f)$ in QIE, and it is known after $\beta(t)$ is solved.
The obtained $\beta(t_f)$ (uncontrolled but known) is then adapted in the target state in Eq. (\ref{psitar}).
It is this freedom in $\beta(t_f)$ that allows us to satisfy the quasiadiabatic condition for any given $c$ value, while it is a fixed value in FAQUAD. 
With known $\beta(t_f)$'s (as shown in Table \ref{table1} for different $c$ values), we can obtain any desired relative phase in the target superposition states by adding a phase to the Rabi frequency $\Omega(t)$.
 
\begin{table}[b]%The best place to locate the table environment is directly after its first reference in text
\caption{\label{table1}%
The $c$ values used to solve for the robust pulses. The corresponding $\beta(t_f)$ and pulse area of the solutions are shown in the second and third columns.
}
%\begin{ruledtabular}
\begin{indented}
\lineup
\item[]\begin{tabular}{ccc}
\br  
c&
$\beta(t_f) (\times\pi)$&
\textrm{Area} $(\times\pi)$\\
\mr
0.073 & 0.051 & 1.970\\
0.060 & 0.034 & 2.470\\
0.050 & 0.033 & 3.076\\
0.040 & 0.023 & 3.839\\
\br
\end{tabular}
\end{indented}
%\end{ruledtabular}
\end{table}

For a small $c=0.073$ (giving a pulse area of 1.97$\pi$ which is equal to the third-order SSSP in Ref. \cite{ndong15}), we solve for $\beta(t)$ and show the result in Fig. \ref{f1}(a). 
The corresponding $\Omega(t)$ and $\Delta(t)$ are shown in Figs. \ref{f1}(b) and \ref{f1}(c), respectively.
In Fig. 1(d), we show the calculated adiabaticity parameter $\left|\frac{\dot{\Omega}\Delta-\Omega\dot{\Delta}}{2(\Omega^2+\Delta^2)^{3/2}}\right|$ using the derived pulse parameters. 
Different from conventional adiabatic schemes, the adiabaticity parameter is a constant throughout the process in QIE, verifying that quasiadiabaticity is indeed achieved.

\begin{figure}[b]
\center
\includegraphics[width=4in]{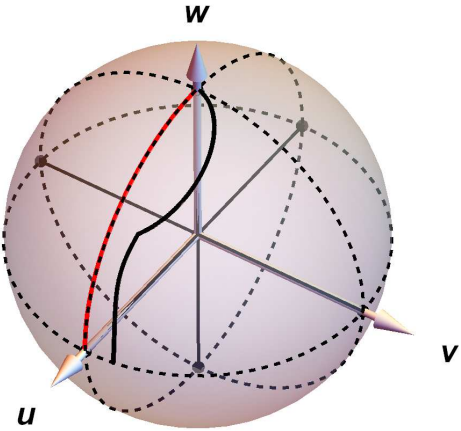}%
\caption{\label{fbloch} Bloch vector motion of the coherent superposition designed by QIE ($c$=0.073). The red arc from the $w$ axis to the $u$ axis represents the adiabatic state.} 
\end{figure}

The time evolutions of the populations in the basis $|1\rangle$ and $|2\rangle$ are shown in Fig. \ref{f2}(a).
Clearly, the system evolves to the desired coherent superposition of states as required by the boundary conditions.
QIE ensures high fidelity transfer, which is not guaranteed in conventional adiabatic processes.
The time evolutions in the adiabatic basis are shown in Fig. \ref{f2}(b).
As required by quasiadiabaticity, the population follows the adiabatic states closely, which is different from the population switching resulting from a phase jump observed in the SSSP approach \cite{ndong15}.
QIE provides a method to design system evolution with a fixed adiabaticity parameter and ensures high fidelity transition.
By defining the Bloch variables $u=\sin\theta(t)\cos\beta(t)$, $v=\sin\theta(t)\sin\beta(t)$, $w=\cos\theta(t)$ \cite{vitanov15}, we show the corresponding Bloch vector motion in Fig. \ref{fbloch}.
We can observe that the Bloch vector stays close to the adiabatic state (red arc) with a small $v(t)$ value throughout the process, which is in contrast to the adiabatic-like evolutions in ref. \cite{vitanov15} where $v(t)$ can be close to 0.5.

\begin{figure}[b]
\center
\includegraphics[width=4in]{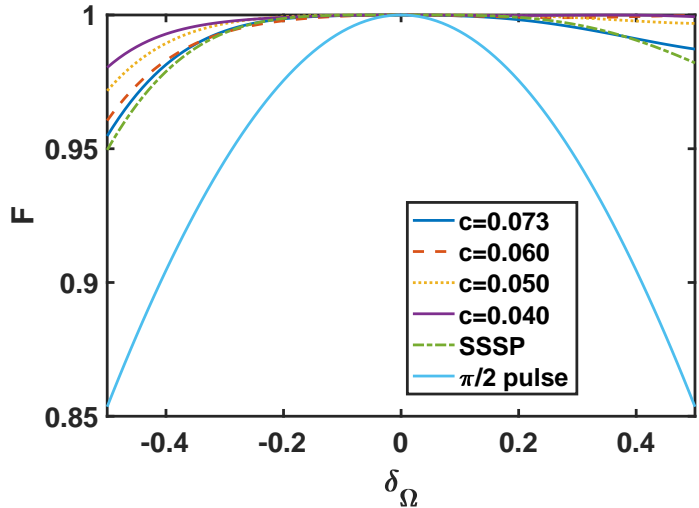}%
\caption{\label{f3} Comparison of the robustness against systematic error in Rabi frequency for different $c$ values using QIE. The robustness of the third-order SSSP \cite{ndong15} and $\pi/2$-pulse are also shown for comparison.}
\end{figure}

\begin{figure}
\center
\includegraphics[width=4in]{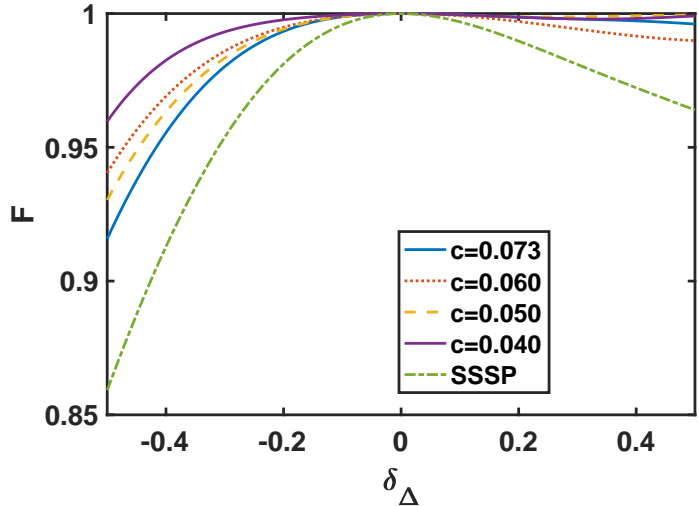}%
\caption{\label{f4} Comparison of the robustness against systematic error in detuning for different $c$ values using QIE. The robustness of the third-order SSSP \cite{ndong15} is also shown for comparison.}
\end{figure}

\subsection{Robustness against errors}
Next, we analyze the robustness with respect to errors in Rabi frequency $\Omega(t)$ and detuning $\Delta(t)$. 
We define fidelity $F$ as the overlap between the final state and the target state $F=|\langle\Psi(t_f)|\Psi_{tar}\rangle|^2$.
We first consider systematic error in Rabi frequency in the form of $\delta_\Omega\Omega(t)\sigma_x/2$, where $\delta_\Omega$ is the amplitude of the relative error.
The robustness with respect to $\delta_\Omega$ for different values of $c$ is shown in Fig. \ref{f3}.
Table \ref{table1} summarizes the $c$ values with their corresponding $\beta(t_f)$'s and pulse areas.
The excitation profiles of the simple $\pi/2$-pulse and the third-order SSSP \cite{ndong15} are also shown in the same figure.
As expected, these pulses show better excitation profiles than the $\pi/2$-pulse.
The SSSP protocol is optimized against pulse area variations using perturbation theory, and it has a flatter excitation profile than the $c=0.073$ protocol (same pulse area of 1.97$\pi$) for small $\delta_\Omega$.
For larger $\delta_\Omega$, we can see that the $c=0.073$ protocol has a small advantage over the SSSP protocol.
For smaller $c$ values, the system is driven closer to adiabaticity, and we observe that the transfer profile is flatter.
At the same time, we can see from Table \ref{table1} that the pulse area increases as $c$ is reduced.

Next, we consider systematic error in detuning in the form of $-\delta_\Delta\Delta(t)\sigma_z/2$, where $\delta_\Delta$ is the amplitude of the relative error.
The robustness with respect to $\delta_\Delta$ for different values of $c$ is shown in Fig. \ref{f4}.
The excitation profile of the third-order SSSP is also shown in the same figure.
The SSSP is optimized against pulse area variations only \cite{ndong15}, so it is not robust against errors in detuning.
Again, as the $c$ value is reduced, the transfer profile is flatter.
We can observe that the proposed protocols are robust against errors in both Rabi frequency and detuning, a result of driving the system as close to adiabaticity as possible.

\section{DISCUSSION AND CONCLUSION}
In inverse engineering techniques, the desired state evolution is designed first, and then the Hamiltonian is obtained.
Without imposing limitations on the Hamiltonian as Eq. (\ref{faquad2}), system adiabaticity cannot be controlled.
It was first proposed in Ref. \cite{vitanov15} that inverse engineering techniques can be used to design adiabatic-like quantum state evolutions, but the adiabaticity and robustness of these adiabatic-like protocols are not analyzed.
In our approach, we use a single control parameter $\beta(t)$ to drive the system as close to adiabaticity as allowed by the available energy (pulse area, determined by the choice of $c$) for any chosen $T$, and the robustness against errors in Rabi frequency and detuning results naturally from quasiadiabaticity. 
We note that it is possible to design SSSP protocols that are robust against errors in Rabi frequency and detuning simultaneously using  the global phase as a single control parameter to nullify the error integrals \cite{daems13}. 
What we show here is a simple alternative approach to obtain robustness in Rabi frequency and detuning simultaneously without the need to nullifying integrals, and the dynamics of the system evolution is very different from the SSSP technique as shown in Fig. \ref{f2}(b).
Similar strategy has been developed to optimize adiabaticity in coupled-waveguide devices \cite{ho15}, where inverse engineering is used to engineer the system evolution to be as close to the adiabatic state as possible.
The major difference is that, instead of the constant adiabaticity criterion in this work, the phase parameter $\beta(t)$ in Ref. \cite{ho15} is expanded using Fourier series to approximate a target phase function.
The adiabaticity parameter still fluctuates during the process in that protocol, and from the discussions in Ref. \cite{sofia15}, the quasiadiabaticity (constant adiabaticity) condition should provide a faster shortcut to adiabaticity.
Compared with other local adiabaticity strategies \cite{guerin02,guerin11,quan10,roland02,sofia15}, the current strategy provides the exact target state at short process times through inverse engineering.
The STA strategies \cite{ruschhaupt12,lu13,daems13,ndong15} can provide exact target state as well as robustness against errors by design, while  the current approach offers robustness by driving the system as close to adiabaticity as possible, providing a versatile alternative to the aforementioned manipulation techniques.
As discussed in Refs. \cite{ruschhaupt12,lu13}, ``robustness'' is a relative concept depending on the kind of noise and perturbations.
A protocol may be robust with respect to certain errors but not to others.
The approach we put forward here is robust in the sense that the insensitivity to fluctuations and uncertainty are provided by adiabaticity.
We also note that the derived pulses under the set conditions are not unique.
Various solutions exist for the system of nonlinear differential equations, and suitable solutions must be chosen depending on the physical constraints (e.g. maximum pulse amplitude, pulse area, etc.).

We have put forward a robust protocol for coherent population or state control of a quantum system using quasiadiabatic inverse engineering.
The inverse engineering STA approach generates the exact target state, and quasiadiabaticity ensures that the evolution follows the adiabatic states as close as possible.
Due to the quasiadaibatic nature of the transfer, the protocol is robust against errors in Rabi frequency and detuning.
Although the current work only focuses on the preparation of coherent superposition of states, we note that the QIE is a general coherent control protocol that can also be used in population inversion/transfer.
This technique can also find applications in integrated optics \cite{longhi09,tseng14,sofia17,chung17}, nonlinear frequency conversion \cite{suchowski14}, and polarization conversion \cite{wen14}.

\section*{Acknowledgments}
We thank J. G. Muga for discussions and comments. 
This work was supported in part by the Ministry of Science and Technology (MOST) of Taiwan (Grant No. 105-2221-E-006-151-MY3, 106-3113-M-110-001).
%\end{acknowledgments}

\section*{References}
%\bibliography{v1.bib}
%\bibliographystyle{unsrt}


\end{document}